\documentclass[fleqn,twoside]{article}
\usepackage{espcrc2}
\usepackage{graphicx}
\usepackage{psfrag}
\usepackage{amsmath}

   \newcommand{\no}{\nonumber}
\newcommand{\nn}{\nonumber \\} \newcommand{\ov}[1]{\overline{#1}}
\newcommand{\eq}[1]{Eq.~(\ref{#1})}
\newcommand{\eqsand}[2]{Eqs.~(\ref{#1}) and (\ref{#2})}

\newcommand{\gev}{\,\mbox{GeV}} \newcommand{\mev}{\,\mbox{MeV}}

\newcommand{\gsim}{\stackrel{>}{_\sim}}

\newcommand{\BDTN}{B \to D \tau \nu_{\tau}}

\newcommand{\BDLN}{B \to D \ell \nu_\ell}

\newcommand{\BTN}{B \to \tau \nu_\tau}
\newcommand{\BTNp}{B^+ \to \tau^+ \nu_\tau}

\newcommand{\fig}[1]{Fig.~\ref{#1}}

\title{\vspace{-1.8cm}{\small TTP08-33\hfill SFB/CPP-08-54}\\[1cm] Higgs
  hunting with B decays\thanks{Talk at \emph{Second Workshop on Theory,
      Phenomenology and Experiments in Heavy Flavour Physics}, June
    16-18 2008, Anacapri, Italy.  Work supported by DFG grant NI
    1105/1--1, SFB-TR09 and the EU Contract MRTN-CT-2006-035482, \lq\lq
    FLAVIAnet''.}}

\author{Ulrich Nierste\address{
        Institut f\"ur Theoretische Teilchenphysik (TTP)\\ 
        Karlsruhe Institute of Technology (Universit\"at Karlsruhe),\\         
        76131 Karlsruhe, Germany}}%

\begin{document}
\thispagestyle{empty}
\begin{abstract}
  B physics is sensitive to the effects of Higgs bosons in the 
  Minimal Supersymmetric Standard Model, if 
  the parameter $\tan\beta$ is large. I briefly summarise the role 
  of $B\to \mu^+\mu^-$ and $\BTNp$ in the hunt for new Higgs effects 
  and present new results on the decay $\BDTN$:  
  Using the analyticity properties of form factors one can predict the
  ratio $R\equiv\mathcal{B}(\BDTN)/\mathcal{B}(\BDLN)$, $\ell=e,\mu$,
  with small hadronic uncertainties. In the Standard Model one
  finds $R= 0.31 \pm 0.02$, ${\cal B} (B^- \to D^0 \tau^-
  \bar{\nu}_{\tau}) = (0.71\pm 0.09)\% $ and ${\cal B} (\bar{B}^0 \to
  D^+ \tau^- \bar{\nu}_{\tau})= (0.66\pm 0.08)\% $, if the vector form
  factor of the Heavy Flavor Averaging Group is used. $\BDTN$ is
  competitive with $\BTNp$ in the search for effects of charged Higgs
  bosons. Especially sensitive to the latter is the differential
  distribution in the decay chain $\bar{B}\to
  D\bar{\nu}_{\tau}\tau^-[\to\pi^-\nu_{\tau}]$.  
\end{abstract}

\maketitle

\boldmath
\section{Higgs effects in B physics}
\unboldmath%
Weakly-coupled extensions of the Standard Model (SM) typically possess a
richer Higgs sector than the latter. The easiest extension of the SM
Higgs sector involves one additional Higgs doublet and is realised in
the Minimal Supersymmetric Standard Model (MSSM). At tree-level the MSSM
Higgs sector coincides with a Two-Higgs-doublet model (2HDM) of type II,
in which down-type fermions receive their masses solely from one doublet,
while up-type fermion masses exclusively stem from Yukawa interactions
with the other Higgs doublet. An important parameter is the ratio
$\tan\beta$ of the two vacuum expectation values.  In the type-II 2HDM
the bottom and top Yukawa couplings $y_b$ and $y_t$ satisfy the relation
\begin{eqnarray}
\frac{y_b}{y_t} &=& \frac{m_b}{m_t} \tan\beta . \no
\end{eqnarray} 
Values around $\tan\beta={\cal O}(60)$ correspond to $y_b$--$y_t$
unification, which occurs in grand-unified theories (GUTs) with a
minimal Yukawa sector.  The idea of grand unification seems to call for
low-energy supersymmetry, which stabilises the electroweak scale against
radiative corrections from heavy GUT particles, improves the unification
of the gauge couplings and reconciles the prediction of the proton
lifetime with its experimental bounds.  Probing the large-$\tan\beta$
region of the MSSM is therefore of great interest, since the question of
Yukawa unification sheds light on the Yukawa sector of the underlying
GUT theory. Yet large values of $\tan\beta$ are also interesting from
purely phenomenological considerations: The tension between the measured
anomalous magnetic moment of the muon, $a_\mu$ \cite{BNL6}, and the
Standard Model prediction \cite{reviews} invites supersymmetry with
$\tan\beta \gsim 10$, and larger values of $\tan\beta$ allow to saturate
$a_\mu$ with heavier superpartners.  Recent global fits of electroweak
and B-physics observables to the constrained MSSM and the model with
minimal gauge-mediated supersymmetry breaking gave best fits for values
of $\tan\beta=54$ and $\tan\beta=55$, respectively \cite{hmsw}.

B physics is excellently suited to study large-$\tan\beta$ scenarios,
because down-type Yukawa couplings grow with $\tan\beta$ and
$\tan\beta={\cal O}(50)$ corresponds to $y_b\sim 1$ \cite{btanb}.
Most dramatic effects can be expected in the leptonic decays $B_q\to
\ell^+\ell^-$ (with $q=d$ or $s$ and $\ell=e,\mu$ or $\tau$), which
are not only loop-suppressed in the Standard Model but also suffer from
an additional helicity suppression. In particular the 95\% CL limit
\begin{equation}
\begin{split}
{\cal B} (B_s\to \mu^+\mu^-) \,\leq &\, 5.8 \cdot 10^{-8} \nn 
 & \approx \;
  18 \cdot {\cal B}^{\rm SM} (B_s \to \mu^+\mu^-)~~~
\no
\end{split}
\end{equation}
from the CDF experiment \cite{bmumu} already cuts into the
large-$\tan\beta$ region of the MSSM parameter space \cite{btanb}. This
is even true for the popular scenario of Minimal Flavour Violation (MFV)
\cite{dgis}, in which the supersymmetric contribution to the $B_s\to
\mu^+\mu^-$ amplitude involves the same elements of the
Cabibbo-Kobayashi-Maskawa (CKM) matrix as the SM amplitude. 
For large $\tan\beta$ one finds  
\begin{eqnarray}
{\cal B} (B_s \to \mu^+\mu^-) & \propto&
     \epsilon_Y^2 \frac{\tan^6 \beta}{M_{A^0}^4},
\label{bmumususy}
\end{eqnarray}
where $M_{A^0}$ is the mass of the CP-odd Higgs boson and $\epsilon_Y$
is a loop function which depends on several MSSM parameters
\cite{btanb}. The pattern of \eq{bmumususy} is in sharp contrast with
the case of the naive type-II 2HDM, in which ${\cal B} (B_s \to
\mu^+\mu^-)$ is proportional to only four powers of $\tan\beta$
\cite{ln}. In the LHCb experiment it will be possible to measure ${\cal
  B} (B_s \to \mu^+\mu^-)$ for any value of $\tan\beta$. One can then
test the MFV hypothesis by checking whether ${\cal B} (B_d \to
\mu^+\mu^-)/{\cal B} (B_s \to \mu^+\mu^-)$ agrees with
$|V_{td}/V_{ts}|^2 f_{B_d}^2/f_{B_s}^2$, where $f_{B_q}$ is the decay
constant of the $B_q$ meson.

Due to the overall loop factor of $\epsilon_Y$  the
bound on $\tan^3\beta/M_{A^0}^2$ derived from \eq{bmumususy} depends on a
plethora of other MSSM parameters. By contrast, effects of the charged
Higgs boson $H^+$ enter $B$ decays at the tree-level. The information
gained from charged-Higgs-mediated processes is therefore more directly
related to the parameters of the MSSM Higgs sector, with smaller
dependences on e.g.\ superpartner masses. $H^+$ effects are best studied
in leptonic and semi-leptonic $B$ decays, in which hadronic uncertainties
are under sufficient control. We specify our discussion to the case of a
$\tau $ lepton in the final state, because the Yukawa couplings of the
third fermion generation are largest. The B factories have observed
the decay $B\to \tau \nu$ with \cite{bpexp} 
\begin{eqnarray}
{\cal B} (B\to \tau \nu ) &=& (1.41\pm 0.43)\times10^{-4} 
 ,  \label{btnexp}  
\end{eqnarray}
which allows to place first useful constraints on $\tan\beta/M_{H^+}$
\cite{hip}. In the following sections I will elaborate on another
promising charged-Higgs hunting ground, the decay $B\to D \tau \nu$, and
compare this mode with $\BTN$. The presented work has been performed in
collaboration with St\'ephanie Trine and Susanne Westhoff \cite{ntw}.

\boldmath
\section{Charged-Higgs effects at large $\tan\beta$}
\unboldmath%
The $ \ov{q} b H^+$ coupling (with $q=u$ or $c$) is given
by
\begin{eqnarray}
{\cal L}_{ \ov{q} b H^+} =  - \frac{g}{2\sqrt{2}} \frac{\ov m_b}{M_W}
      \frac{\tan\beta}{1+\epsilon_b \tan\beta}
      \, V_{qb} \cdot \nn
      \qquad \ov{q} (1+\gamma_5) b H^+ ,
\label{lbqh}
\end{eqnarray}
where the small Yukawa coupling $y_q$ is set to zero.  The bottom quark
mass $\ov m_b$ is defined in the same QCD renormalisation scheme as the
current $\ov q (1+\gamma_5) b$. In the 2HDM of type-II the parameter
$\epsilon_b$ vanishes. In the MSSM with MFV $\epsilon_b$ is a loop
factor; the typically dominant squark-gluino contribution to
$\epsilon_b$ is proportional to $\mu^*/M_{\tilde g}$, where $\mu$ is the
Higgsino mass parameter and $M_{\tilde g}$ is the gluino mass
\cite{btanb}. A priori $\epsilon_b$ could be complex, but experimental
constraints from electric dipole moments severely constrain the phase of
$\mu^*/M_{\tilde g}$ and thereby of $\epsilon_b$. Since $|\epsilon_b|
\tan\beta$ can be of order 1, the charged-Higgs phenomenology does
involve genuine supersymmetric parameters, yet with much less impact
than in $B_s \to \mu^+\mu^-$.  In the MSSM with a generic flavour
structure $\epsilon_b$ is different for $q=u$ and $q=c$ and may obtain a
sizable phase. In the generic 2HDM $\epsilon_b$ is generated at
tree-level and is typically complex. For an early extensive study of
$\BDLN$ and $\BDTN$ in the MFV-MSSM see Ref.~\cite{iko}.

\begin{figure}[t]
{\psfrag{ix}{\raisebox{0.1cm}{\hspace{-0.2cm}$g_P$}}
\psfrag{iy}{\raisebox{0cm}{\hspace{-1.2cm}
$\mathcal{B}(B\to\tau\nu)\,[10^{-3}]$}}
\psfrag{X1}{$1\sigma$}
\psfrag{X2}{$3\sigma$}
\centerline{\includegraphics[scale=0.95]{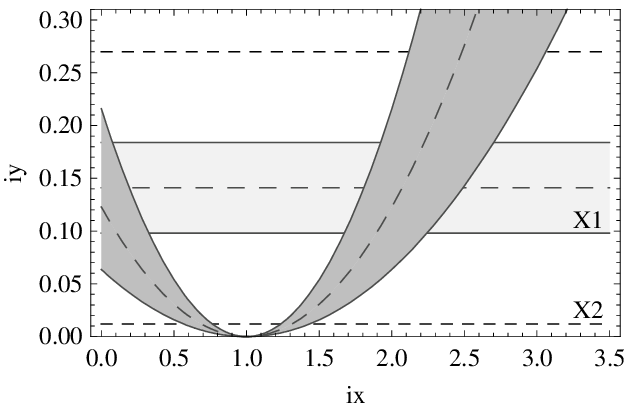}}
}\vspace{-5ex}
\caption{${\cal B}(B^+ \to \tau^+ \nu)$ vs.\ $g_P$ for real $g_P$.
\label{fig:bpgp}}
\end{figure}

The effective hamiltonian describing $b\to q \tau \nu$ transitions
mediated by $W^+$ or $H^+$ 
can be written as
\begin{eqnarray}
H_{\textrm{eff}} &\!\!=\!\!& 
\frac{G_{F}}{\sqrt{2}} V_{qb}\text{\thinspace}
\big\{ \overline{q}\gamma^{\mu}(1-\gamma_{5})b \text{\thinspace}
       \overline{\tau}\gamma_{\mu}(1-\gamma_{5})\nu_{\tau} \nn
&& \; -\frac{\ov m_{b}m_{\tau}}{m_{B}^{2}}\text{\thinspace}\text{\thinspace}
\overline{q}\left[ g_{S} + g_P \gamma_5\right] b  \text{\thinspace}
\overline{\tau}(1-\gamma_{5})\nu_{\tau} \big\}\nn
&& \,+\,\mbox{h.c.} \label{eq:1}  
\end{eqnarray}
In the MSSM the effective couplings $g_S$ and $g_P$ read
\begin{equation}
 g_{S}\! =\! g_{P}\! =\! \frac{m_{B}^{2}}{M_{H^+}^{2}} 
\frac{\tan^{2}\beta}{ (1+\epsilon_b\,\tan\beta)
                      (1+\epsilon_{\tau}\,\tan\beta)} .
\label{gsgp}
\end{equation}
Here $\epsilon_{\tau}$ is the analogue of $\epsilon_b$ for the 
$ \ov{\nu}_\tau \tau H^+$ coupling. $B^+ \to \tau^+ \nu$ probes the
coupling $g_P$ \cite{hip}: 
\begin{equation}
{\cal B}(B^+ \to \tau^+ \nu) \; \propto \;
     |V_{ub}|^2 f_B^2 \, |1- g_P | ^2 \label{b+tn}
\end{equation}
The $B$ meson decay constant $f_B=216\pm 38\,\mev$ \cite{latt07} is the
dominant source of theoretical uncertainty in the extraction of $|1-
g_P |$ from ${\cal B}(B^+ \to \tau^+ \nu)$. \eq{b+tn} is confronted with
the experimental result of \eq{btnexp} in \fig{fig:bpgp}.

The semi-tauonic decay $B\to D\tau \nu_\tau$ instead probes the coupling
$g_S$, because $B$ and $D$ have the same parity. In the MFV-MSSM $B\to
D\tau \nu_\tau$ and $B^+ \to \tau^+ \nu$ probe the same parameters,
because $\epsilon_b$ in \eqsand{lbqh}{gsgp} is the same for $b\to u$ and
$b\to c$ transitions. Thus within the MFV-MSSM we can combine the
information from both decay modes to constrain the combination of
$\tan\beta$, $M_{H^+}$, $\epsilon_b$ and $\epsilon_{\tau}$ in \eq{gsgp}.
If new physics is found, a comparison of $g_S$ extracted from $B\to
D\tau \nu_\tau$ with $g_P$ obtained from $B^+ \to \tau^+ \nu$ will probe
physics beyond the MFV-MSSM. 
$B\to D \tau \nu$ compares to $B^+ \to \tau^+ \nu$ as follows: 
\begin{itemize}
\renewcommand{\itemsep}{-2pt}
 \item[i)]  $ {\cal B} (\BDTN) \approx 50\, {\cal B} (\BTNp)$.
 \item[ii)] 
    $ |V_{cb}|$ entering $ \BDTN$ is much better known than
    $ |V_{ub}|$.
  \item[iii)] The uncertainty in lattice calculations of $ f_B^2$ needed
    for $\BTN$ is 30\%. Instead $ B\to D \tau \nu$ involves hadronic
    form factors. We will see below in Sect.~\ref{sect:bdtn} that the
    associated theoretical uncertainty is smaller, if experimental data
    on $ \BDLN$ with $\ell=e,\mu$ are exploited.
 \item[iv)] The three--body decay
    $ \BDTN$ has decay distributions
    which discriminate between $ W^+$ and $ H^+$ exchange.
 \item[v)]  $ \BTN$ is mildly helicity--suppressed.
    In $ \BDTN$ the (transverse)
    $ W^+$ contribution is helicity suppressed
    in the kinematic region with slow $ D$ meson
    (P--wave suppression) \cite{gh}:
\mbox{
\hspace{0ex}\includegraphics[scale=.38, angle=0, clip=false]{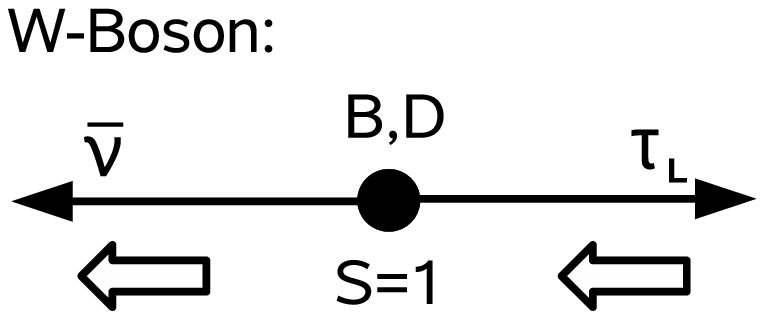}
~\includegraphics[scale=.38, angle=0, clip=false]{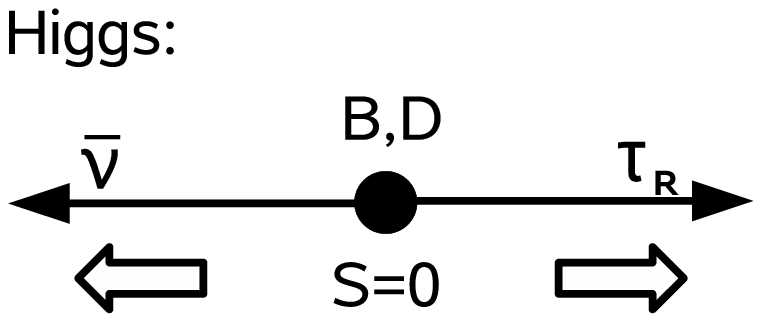}}
\end{itemize}

\section{Charged-Higgs effects in $\BDTN$}\label{sect:bdtn} 
\begin{figure}[t!]
  {\psfrag{ix}{\raisebox{0.05cm}{\hspace{-0.3cm}\scalebox{1}{$w$}}}
    \psfrag{iy}{\raisebox{0cm}{\hspace{-0.5cm}
      \scalebox{0.7}{$|V_{cb}|V_1(w)$}}}
    \psfrag{1.0}{\hspace{-0.05cm}$\scalebox{0.8}{1.0}$}
    \psfrag{1.1}{\hspace{-0.05cm}$
    \scalebox{0.8}{1.1}$}\psfrag{1.2}{\hspace{-0.05cm}$\scalebox{0.8}{1.2}$}
    \psfrag{1.3}{\hspace{-0.05cm}$\scalebox{0.8}{1.3}$}
    \psfrag{1.4}{\hspace{-0.05cm}$\scalebox{0.8}{1.4}$}
    \psfrag{1.5}{\hspace{-0.05cm}$\scalebox{0.8}{1.5}$}
    \psfrag{1.6}{\hspace{-0.05cm}$\scalebox{0.8}{1.6}$}
    \psfrag{0.02}{\hspace{-0.09cm}$\scalebox{0.8}{0.02}$}
    \psfrag{0.03}{\hspace{-0.09cm}$\scalebox{0.8}{0.03}$}
    \psfrag{0.04}{\hspace{-0.09cm}$\scalebox{0.8}{0.04}$}
    \psfrag{0.05}{\hspace{-0.09cm}$\scalebox{0.8}{0.05}$}
    \psfrag{0.06}{\hspace{-0.09cm}$\scalebox{0.8}{0.06}$}
    \psfrag{0.07}{\hspace{-0.09cm}$\scalebox{0.8}{0.07}$} \hspace{-0.4cm}
    \centerline{\includegraphics[scale=0.82]{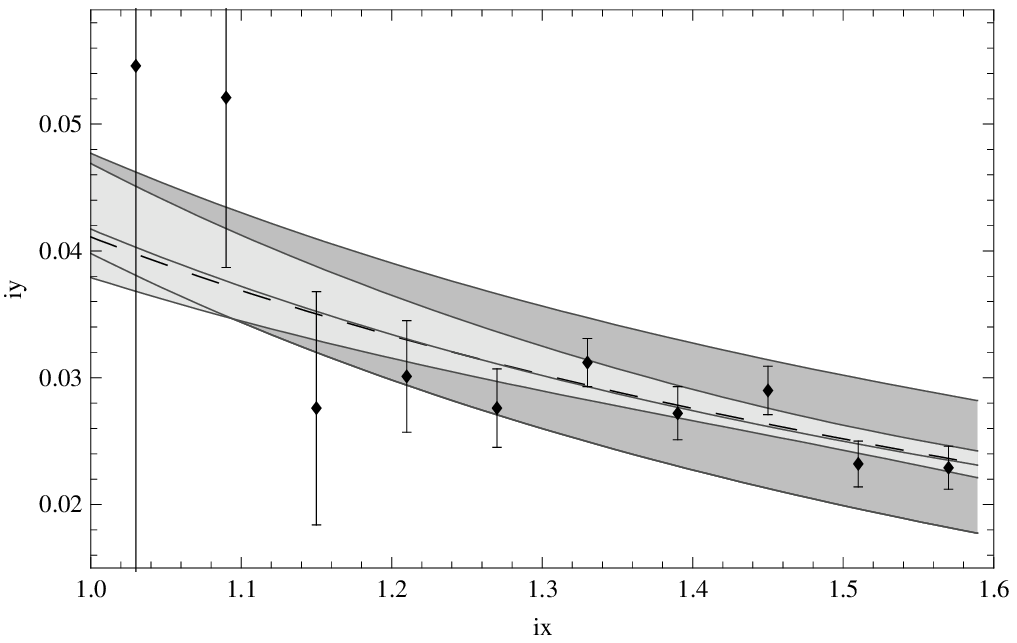}} }
  \vspace{-1.0cm}
\caption{Vector form factor $V_1(w)$. Dots: BELLE data on $\BDLN$
  \cite{b01} with statistical errors only. Dark gray band: form factor
  from BELLE data with HQET constraint at $w=1$ (systematic errors
  dominate at large recoil). Light gray band: form factor from HFAG, 
  which includes older CLEO and ATLAS data 
  \cite{HFAG}. Solid line: best fit of BELLE data to $a_0^V$ and
  $a_1^V$. Dashed line: best fit of Ref.~\cite{b01} (using different
  parameters).}
\label{fig:v1}
\end{figure}

\renewcommand{\thefootnote}{\fnsymbol{footnote}}
$\BDLN$, $\ell=e,\mu,\tau$, involves the hadronic matrix elements of the
vector current and of the scalar current, which are expressed in terms
of two form factors,\footnote{There is a typo in the definition of $S_1$
in Ref.~\cite{ntw}.} $V_1$ and $S_1$:
\begin{eqnarray}
\lefteqn{\langle D(p_D)|\bar{c}\gamma^{\mu}b|\bar{B}(p_B)\rangle
  =}\; && \nn
&& V_1(w) \frac{m_B+m_D}{2 \sqrt{m_B m_D}}  
\! 
  \left[ p_B^{\mu}+p_D^{\mu}-\frac{m_B^2-m_D^2}{q^2}q^{\mu}\right]\nn
&&  +\; S_1(w) \, (1+w) \sqrt{m_B m_D}\,
     \frac{m_B-m_D}{q^2}q^{\mu} \nn
\lefteqn{\langle D(p_D)|\bar{c}b(\mu)|\bar{B}(p_B)\rangle = }\; &\nn
&& S_1(w) \, (1+w) \sqrt{m_B m_D}\,
     \frac{m_B-m_D}{\ov m_b(\mu)-\ov m_c(\mu)},  \label{eq:2}
\end{eqnarray}
where $m_D$ and $m_B$ are the meson masses, $q=p_B-p_D$ is the momentum
transfer  and the running quark masses $\ov m_b$ and $\ov m_c$ are
evaluated at the renormalisation scale $\mu$ at which the scalar current
$\bar{c}b$ is defined. The kinematic variable
\begin{eqnarray}
w &=& \frac{m_B^2+m_D^2 -q^2}{2 m_D m_B} \label{defw}
\end{eqnarray}
is defined in such a way that the kinematic endpoint $q^2=(m_B-m_D)^2$ 
corresponds to $w=1$. Heavy quark symmetry implies \cite{hqet}
\begin{eqnarray}
 S_1(1) \; = \; V_1(1) &=& 1 +{\cal O}(1/m_c, \alpha_s) . \label{hqs}
\end{eqnarray}
For $m_\ell=0$ the other kinematic endpoint is at $q^2=0$ corresponding 
to $w_{\rm max}=(m_B^2+m_D^2)/(2m_D m_B)=1.59$. 
The absence of a pole at $q^2=0$ in \eq{eq:2} implies 
\begin{eqnarray}
  S_1(w_{\rm max}) & = &  V_1(w_{\rm max})
. \label{lrc}
\end{eqnarray}
The next step towards a precision analysis of $\BDTN$ exploits the 
analyticity properties of form factors \cite{analy}: The location 
of poles and branch points in $V_1$ and $S_1$ can be inferred from the 
particle spectrum. The conformal mapping 
\begin{displaymath}
 z = \frac{\sqrt{(m_B+m_D)^2 - q^2}- \sqrt{(m_B+m_D)^2 - t_0} }{
        \sqrt{(m_B+m_D)^2 - q^2} + \sqrt{(m_B+m_D)^2 - t_0} }
\end{displaymath}
and the elimination of subthreshold poles renders the form factors
analytic in the new variable $z$ in the entire kinematic region (see
\cite{analy,ntw} for details).  That is, we can paramaterise $V_1$ and
$S_1$ in terms of power series in $z$. With a proper choice of the free
parameter $t_0$ the kinematic range $1\leq w\leq 1.59$ is mapped onto
$0\leq |z|\leq 0.032$. $\BDLN$ with $\ell=e,\mu$ only involves $V_1$. We
can use experimental data to verify that the first two coefficients
$a_0^V$ and $a_1^V$ of the power series are sufficient to describe the
normalisation and shape of $V_1$.  Moreover, the dependence on $a_1^V$
is moderate.  The result is shown in \fig{fig:v1}.
Finally we need the scalar form factor $S(w)$. The corresponding
parameters $a_0^S$ and $a_1^S$ are fixed through \eqsand{hqs}{lrc},
using the $1/m_c$ and $\alpha_s$ corrections from \cite{n}. An
alternative approach uses lattice data to fix the form factors near
$w=1$ \cite{km}. Note that our analysis faces much smaller hadronic
uncertainties than the extraction of $|V_{cb}|$ from $\BDLN$: We first
fix $|V_{cb}| V_1(w)$ from experiment. Then \eq{lrc} determines the
normalisation of $|V_{cb}| S_1(w)$ in terms of measured quantities.
Keeping only $a_0^S$ already reproduces $S(w)$ to 90\%, and neither
hadronic uncertainties nor the error in $|V_{cb}|$ have entered the
prediction of $\BDTN$ yet.  Hadronic uncertainties  only enter when we
increase the accuracy further and fix the slope parameter $a_1^S$ by
including \eq{hqs}. These uncertainities are suppressed by $1/m_c$. In
this step one also needs the experimental value of $|V_{cb}|$, whose
error is around 2\%.  A remaining source of hadronic uncertainty is the
next term $a_2^S$ of the series in $z$.  In the data for $V_1$ in
\fig{fig:v1} we found the influence of $a_2^V$ negligible\footnote{For a
  lattice study of $a_2^V$ in the $B\to\pi$ form factor, which involves
  the much larger range $0\leq |z|\leq 0.28$ than our $B\to D$
  transition, see Paul Mackenzie's talk at this conference.}  and expect
the same behaviour for $a_2^S$ and $S_1$.

\begin{figure}[t]
{\psfrag{ix}{\raisebox{0.1cm}{$g_S$}}
\psfrag{iy}{\raisebox{-0.2cm}{\hspace{-0.3cm}$R$}}
\psfrag{X1}{$1\sigma$}
\psfrag{0.15}{}\psfrag{0.25}{}\psfrag{0.35}{}\psfrag{0.45}{}
\psfrag{0.0}{\hspace{0.06cm}$\scalebox{0.8}{0}$}
\psfrag{0.5}{\hspace{-0.04cm}$\scalebox{0.8}{0.5}$}
\psfrag{1.0}{\hspace{0.06cm}$\scalebox{0.8}{1}$}
\psfrag{1.5}{\hspace{-0.04cm}$\scalebox{0.8}{1.5}$}
\psfrag{2.0}{\hspace{0.06cm}$\scalebox{0.8}{2}$}
\psfrag{2.5}{\hspace{-0.04cm}$\scalebox{0.8}{2.5}$}
\psfrag{3.0}{\hspace{0.06cm}$\scalebox{0.8}{3}$}
\psfrag{3.5}{\hspace{-0.04cm}$\scalebox{0.8}{3.5}$}
\psfrag{0.10}{\hspace{-0.08cm}$\scalebox{0.8}{0.1}$}
\psfrag{0.20}{\hspace{-0.08cm}$\scalebox{0.8}{0.2}$}
\psfrag{0.30}{\hspace{-0.08cm}$\scalebox{0.8}{0.3}$}
\psfrag{0.40}{\hspace{-0.08cm}$\scalebox{0.8}{0.4}$}
\hspace{-0.2cm}
\centerline{\includegraphics[scale=0.8]{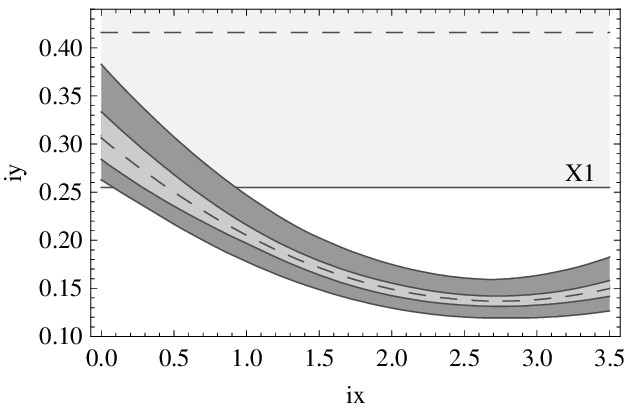}}}
\vspace{-5.5ex}
\caption{$R\equiv\mathcal{B}(\BDTN)/\mathcal{B}(\BDLN)$ as a
  function of $g_S$.  Light gray band:
  $R^{\textrm{exp}}=0.416\pm 0.117\pm 0.052$ \cite{babarbdt}. 
  Dark gray band: $R$ computed with BELLE form factor $V_1$ \cite{b01}.
  Gray band: $R$ computed with HFAG form 
  factor $V_1$ \cite{HFAG}. }\label{fig:br}
\end{figure}

The first application of our analysis in \cite{ntw} is a new prediction of 
branching fractions in the SM: 
\begin{eqnarray}
 {\cal B} (B^- \to D^0 \tau^- \bar{\nu}_{\tau})
   & =& (0.71\pm 0.09)\% \nn
 {\cal B} (\bar{B}^0 \to D^+ \tau^- \bar{\nu}_{\tau})
   & =& (0.66\pm 0.08)\% \qquad  \nn
 R \equiv\frac{\mathcal{B}(\BDTN)}{\mathcal{B}(\BDLN)}
 & =&   0.31 \pm 0.02 \label{bres}
\end{eqnarray}
using the HFAG form factor $ V_1$ \cite{HFAG}. Note the small
uncertainty in $R$ which is the relevant quantity for charged-Higgs
hunting. This has to be compared with the ${\cal O} (40\%)$ error of
$|V_{ub}|^2 f_B^2$ entering the SM prediction of 
$\BTN$. The uncertainties in \eq{bres} will
decrease further with better data on $\BDLN$. 
The dependence of $R$ on $g_S$ is shown in \fig{fig:br}.   
A home-use formula for $R$ as a function of $g_S$ can be found in
Eq.~(7) of Ref.~\cite{ntw}. We notice that the dependence of $R$ on
$g_S$ is weaker than the dependence of ${\cal B}(B^+\to \tau^+ \nu)$ on
$g_P$, but the present $1\sigma$ upper bounds on the effective coupling
constants are similar. However, if nature has opted for a large
charged-Higgs contribution suppressing $R$ to values below 0.2, it will
not be easy to determine $g_S$ because the curve in \fig{fig:br} is
quite flat. At present there is also an open experimental issue in
$\BDTN$: The Monte Carlo simulations use the SM formula for the decay
distribution in $\BDTN$, but the $D$ energy is higher on average if a
charged-Higgs contribution is present, because there is a destructive
interference with the longitudinal $W$ boson contribution. The
efficiencies for $D$ detection are smaller for very soft $D$'s, which
may affect the upper bound on $g_S$ derived from $R^{\textrm{exp}}$.

\begin{figure}[t]
\hspace{-0.2cm}
{\psfrag{ix}{$\cos\theta$}
\psfrag{iy}{\hspace{-2.2cm}$\scriptstyle d\Gamma(B^0\to D^+\pi^-\bar{\nu}\nu)\
  [10^{-16}\textrm{GeV}]$}
\psfrag{X1}{$E_D=2\,\textrm{GeV}$, $E_{\pi}=1\,\textrm{GeV}$}
\psfrag{X2}{$g_S=0$ (SM)}
\psfrag{X3}{$g_S=1+i$}
\psfrag{X4}{$g_S=2$}
\centerline{\includegraphics[scale=0.8]{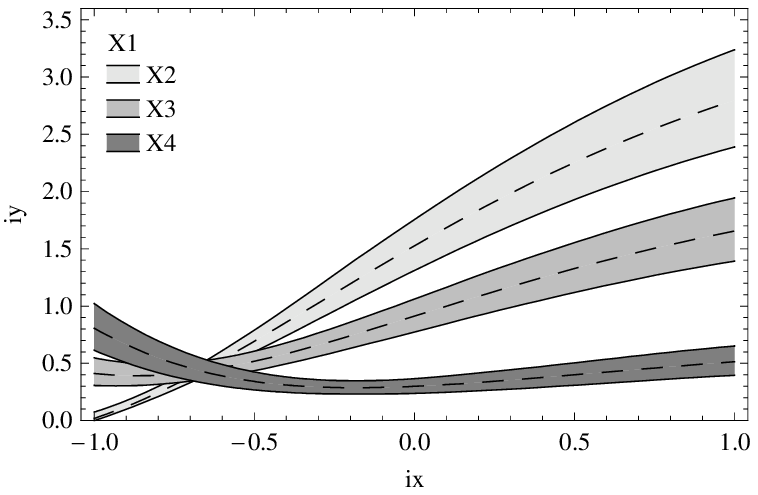}}}
\vspace{-5ex}
\caption{ Angular distribution for $E_D=2\gev$ and $E_{\pi}=1\gev$ and
  $g_S=0,1+i,2$. In models with $g_P=g_S$ these three values correspond
  to the same ${\cal B}(B^+\to \tau^+ \nu)$. }\label{fig:th}
\end{figure}

It is well known that the sensitivity of $\BDTN$ to charged-Higgs
effects is improved, if information on the $\tau$ polarisation is
included \cite{taupol}. However, in $B$ factories the $\tau$ is too slow
to decay with a displaced vertex, so that the $\tau$ kinematics is not
accessible to experiment. To my knowledge, the only theory paper
addressing this problem is Ref.~\cite{ks}, which proposes to study the
differential decay rate $d\Gamma/dE_D$, where $E_D$ is the $D$ meson
energy in the $B$ rest frame. We have studied the decay chain 
$\bar{B}\to D\bar{\nu}_{\tau}\tau^-$ with the subsequent decay 
$\tau^- \to\pi^-\nu_{\tau}$ and propose to study the triple differential
decay rate $d\Gamma/(dE_D\, dE_\pi\, d\cos\theta)$, where $E_\pi$ is the 
$\pi^-$ energy and $\theta$ is the angle between the three-momenta 
of the $D$ and the $\pi^-$. This quantity not only discriminates between 
SM and charged-Higgs effects in an excellent way, it also allows to 
constrain the phase of $g_S$. This feature is illustrated in
\fig{fig:th}.

\section*{Acknowledgements}
I am grateful to St\'ephanie Trine and Susanne Westhoff for an enjoyable
collaboration and thank them for proofreading. Stimulating discussions
with Christoph Schwanda on experimental issues are gratefully
acknowledged.

\end{document}